\documentclass[useAMS]{mn2e}
\usepackage{epsfig,badnell}
\DeclareMathSymbol{\gtrsim}       {\mathrel}{AMSa}{"26}
\DeclareMathSymbol{\lesssim}      {\mathrel}{AMSa}{"2E}
\newcommand{\be}{\begin{equation}}
\newcommand{\ee}{\end{equation}}
\newcommand{\kr}{\kappa_{\rm R}}
\newcommand{\kb}{k_{\rm B}}
\newcommand{\bfx}{{\bf x}}
\newcommand{\rmd}{\,{\rm d}}
\newcommand{\Ne}{N_{\rm e}}
\newcommand{\Ue}{U_{\rm e}}
\newcommand{\me}{{\rm m}_{\rm e}}

\title[Comparison of Rosseland-mean opacities from OP and OPAL]
{A comparison of Rosseland-mean opacities from OP and OPAL}
\author[M. J. Seaton and N. R. Badnell]{M. J. Seaton$^{1}$ and N. R. Badnell$^{2}$\\
$^{1}$Department of Physics and Astronomy, University College London, London WC1E 6BT\\
$^{2}$Department of Physics, University of Strathclyde, Glasgow, G4 0NG}
\begin{document}
\date{Accepted XXX. Received XXX; in original form XXX}

\pagerange{\pageref{firstpage}--\pageref{lastpage}} \pubyear{2004}

\maketitle

\label{firstpage}

\begin{abstract}
Monochromatic opacities from the Opacity Project (OP)
(Seaton et al.) have 
been augmented by hitherto missing inner-shell contributions
(Badnell \& Seaton).
OP  Rosseland-mean opacities, 
$\kr$, are compared with results from OPAL
(Iglesias \& Rogers)
for the elements H, He, C, O, S and Fe. The OPAL data
are obtained from the website
www-phys.llnl.gov/Research/OPAL/index.html.

Agreement for H is close everywhere except for the
region of $\log(T)\simeq 6$ and $\log(R)\simeq -1$
($R=\rho/T_6^3$ where $\rho$ is mass-density in 
g cm$^{-3}$ and $T_6=10^{-6}\times T$ with $T$ in K).
In that region $\kr$(OPAL) is larger than $\kr$(OP)
by up to 13\%. The differences are due to different
equations of state (EOS). In the region concerned
OP has the H ground state undergoing dissolution, 
leading to a small H-neutral ionization fraction, while OPAL has
larger values for that fraction. A similar difference
occurs for He at $\log(R)\simeq-1$ and $\log(T)\simeq 6.4$,
where OP has the He$^+$ ground-state undergoing 
dissolution. 

The OPAL website does not provide single-element
Rosseland means for elements other than H and He. 
Comparisons between OP and OPAL are made for mixtures with $X=0.9$,
$Z=0.1$ and $Z$ containing pure C, O or S. 
There are some differences:
at the lower temperatures, say $\log(T)\leq 5.5$, due to
differences in atomic data, with the OP R-matrix data probably
being the more accurate; and at higher temperatures  mainly
due to differences in level populations resulting from
the use of different EOS theories.

In the original OP work, R-matrix data for iron were
supplemented by data obtained using the 
configuration-interaction (CI) code {\sc superstructure}. 
The experiment is made of replacing much of the 
original iron data with new data from the CI code
{\sc autostructure}. Inclusion of intercombination lines gives an
increase in $\kr$ of up to 18\%.

The OPAL website does not allow for $Z$ containing pure iron.
Comparisons are made for an iron-rich mixture, $X=0.9$,
$Z=0.1$ and $Z$ containing C and Fe with C:Fe=2:1 by
number fraction. There are some differences between OP and OPAL for
that  case: the OP `$Z$-bump' in $\kr$ is shifted to slightly
higher temperatures, compared to OPAL. 

Overall, there is good agreement between OP and OPAL Rosseland-mean
opacities for the 6-elements, but there are some differences. 
Recent work (Bahcall et al.) has shown that 
helioseismology measurements give a very accurate
value for the depth of the solar convection zone, $R_{\rm CZ}$, and
that solar models give agreement with that value only if
opacities at $R_{\rm CZ}$ are about 7\% larger than OPAL values. For
the 6-element mix at $R_{\rm CZ}$ we obtain $\kr$(OP) to be larger
than $\kr$(OPAL) by 5\%.

%\end{quotation}
\end{abstract}
\section{Introduction}
Rosseland-mean opacities from the Opacity Project (OP),
as originally presented in \cite{symp}, were in
good agreement with those from the OPAL project \cite{opal} over much
of the temperature--density plane, but were smaller than those
from OPAL at high temperatures and densities. Iglesias and Rogers 
\cite{IR96} offered the explanation that OP was missing some data
for inner-shell transitions and that was confirmed in a recent
paper by the present authors \cite{bs03}. The results of \cite{IR96}
and \cite{bs03} were for a mixture of 6 elements, H, He, C, O, S and
Fe with abundances, by number-fractions, given in Table 1: we refer
to that as the 6-element mix; mass-fractions are $X=0.7$ for H,
$Y=0.28$ for He, $Z=0.02$ for `metals'.

Fig. 1 shows the level of agreement between OP and OPAL, for
that mix, as obtained
in \cite{bs03}: values of $\log(\kappa_{\rm R})$, where
$\kappa_{\rm R}$ is Rosseland-mean opacity in cm$^2$ g$^{-1}$,
are plotted against $\log(T)$ for 5 different values of
$\log(R)$ where $R=\rho/T_6^3$, $\rho=$  density in g cm$^{-3}$
and $T_6=10^6\times T$ with $T$ in K (fig. 1 of \cite{symp}
shows the behaviour of $\log(R)$ for a few typical stellar
models). The OPAL data are obtained
from the OPAL website \cite{web} and the OP data are with inclusion
of the inner-shell contributions discussed in \cite{bs03}. It is
seen that the agreement between OP and OPAL is fairly good in all cases
but that there are some differences. The purpose of the present
paper is to make more detailed comparisons of OP and OPAL
for the six elements H, He, C, O, S and Fe. Work on the inclusion
of OP inner-shell data for other elements is in progress.
\section{Rosseland means}
Let $\sigma_k(u)$ be the cross-section for absorption or scattering
of radiation by element $k$, where $u=h\nu/(\kb T)$, $\nu$ is 
frequency and $\kb$ is Boltzmann's constant. For a mixture of
elements with number fractions $f_k$, $\sum_kf_k=1$, put 
$\sigma(u)=\sum_kf_k\sigma_k(u)$ (see section \ref{4P1}). The Rosseland-mean cross
section is $\sigma_{\rm R}$ where,
\be \frac{1}{\sigma_{\rm R}}=\int_0^\infty\frac{F(u)}{\sigma(u)}\rmd  u \label{ross}\ee
and 
\be F(u)=\left[\frac{15}{(4\pi^4)}\right]u^4\exp(-u)/[1-\exp(-u)]^2.\label{weight}\ee
The Rosseland-mean opacity per unit mass is $\kr=\sigma_{\rm R}/\mu$
where $\mu$ is mean molecular weight.
\section{Equations of state}
The populations of the energy levels providing absorption or scattering
of radiation are determined by the equation of state, EOS.
OP and OPAL have very different treatments of the problem:
the former use what is referred to as the `chemical picture';
the latter the `physical picture'.
\subsection{The OP EOS}
Let $i$ be equal to the number of bound electrons in ionization
stage $i$. In the OP work the {\em internal partition function} for
stage $i$ is taken to be \cite{eos1}
\be U_i=\sum_j g_{ij}W_{ij}\exp[-E_{ij}/(\kb T)] \label{part} \ee
where $j$ specifies  an energy level, $g_{ij}$ is a statistical
weight, $W_{ij}$ an {\em occupation probability} and $E_{ij}$ is 
the total energy of level $ij$.
Let $\phi_i$ be the fraction of atoms in stage $i$. In the OP
work (\cite{eos1} and \cite{adoc1})
\be \frac{\phi_i}{\phi_{i-1}}=\frac{U_i}{U_{i-1}}\times\frac{\Ne}{\Ue} \ee
where
\be \Ue=2\left\{\frac{\me\kb  T}{2\pi\hbar^2}\right\}^{3/2} \ee
and $\me$ is the electron mass. Allowance for electron degeneracy, 
and other refinements, is discussed in \cite{ghr} and in section
IV(e) of \cite{eos1}.

In the OP work the $W_{ij}$ are calculated using methods described in
\cite{eos1} which are plausible but not rigorous
(some modifications of the treatment in \cite{eos1} are discussed in
\cite{bs03}).
The OP $W_{ij}$ become small for levels which are sufficiently
highly excited (referred to as `level dissolution'), 
which ensures convergence of the summation in (\ref{part}).
In many circumstances the Boltzmann factors in (\ref{part}),
$\exp[-E_{ij}/(\kb T)]$, become small long before there is
a cut-off due to the $W_{ij}$ becoming small: in those circumstances
the calculated opacities are insensitive to the exact 
values of the $W_{ij}$.
\subsection{The OPAL EOS}
The OPAL approach to the EOS problems is based on the many-body
quantum statistical mechanics of partially ionized plasmas
(see \cite{opale} and \cite{RI92} for a summary of later
work). The level populations obtained by OPAL may be expressed
in terms of the $W$ factors of OP. Comparisons of results 
from OPAL and OP are given in \cite{dgh} for H and H$^+$ and
in \cite{IR96} for hydrogenic C. It is found that, compared
with OP, OPAL has larger populations in the more highly excited
states.
  
The case of hydrogenic C was discussed further in \cite{bs03}
where it was found that OPAL
gave $W$ factors to be surprisingly large for
states which had mean volumes larger than the
mean volumes available per particle in the plasma.
\section{Opacities for mixtures}
In principle, the level populations for any one chemical
element depend on the abundances of all other elements
present in a plasma. However, it would not be practicable to
make complete {\em ab initio} calculations of opacities for
every mixture which may be of interest. Some approximations
must therefore be made.
\subsection{The OP approach}\label{4P1}
The OP $W_{ij}$ depend on the ion micro-field which should, of course,
depend on the chemical mixture. The approximation is made of
using a micro-field independent of mixture (in practice, that
for fully-ionized H and He with $X=0.7$, $Y=0.3$). Monochromatic
opacities are then calculated for each chemical element, as
functions of frequency, on a grid of values of $T$ and $N_{\rm e}$
(in practice, usually with intervals of $\delta\log(T)=0.05$
and $\delta\log(N_{\rm e})=0.5$). The monochromatic opacities 
are then simply added together for the calculation of 
Rosseland means for mixtures. In most cases that procedure does not 
lead to significant error, but there are exceptions
(see, for example,  Section (\ref{smooth})).
\subsection{The OPAL approach}
The OPAL work makes use of interpolation procedures based on 
the concept of `corresponding states' (see \cite{corr}).
Rosseland-mean opacities are available from the OPAL website
\cite{web} for `metal' mass-fractions $Z\leq 0.1$. In general
the use of interpolations does not introduce any important
errors (see \cite{corr}) but there may be some significant
errors near the edges of the domains of the tables provided.
\section{Atomic physics}
\subsection{R-matrix calculations}
Most of the atomic data  used in the original OP work, were obtained
using the R-matrix method, RM \cite{adoc2}. For a system containing
$N$ electrons we use the wave-function expansion
%Mike, split onto two lines for MNRAS 2 col format. Nigel
\begin{eqnarray}
\Psi&=&{\cal A}\sum_p \psi_p({\bf x}_1,\cdots,{\bf x}_{N-1})
\times \theta_p({\bf x}_N)\nonumber\\
&+&\sum_m\Phi_m(\bfx_1,\cdots,\bfx_N)\times c_m \label{rm}
\end{eqnarray}
where: ${\cal A}$ is an anti-symmetrisation operator; $\bfx_i$ is a
space-and-spin co-ordinate for electron~$i$; $\psi_p$ is a function
for a state $k$ of the $(N-1)$-electron core (usually calculated
using a configuration-interaction code); $\theta_p$ is an orbital
function for an added electron; $\Phi_m$ is a function of bound-state
type for the $N$-electron system; $c_m$ is a coefficient. In the R-matrix
method the functions $\theta_k$ and coefficients $c_m$ are fully
optimised.

Further details on the R-matrix calculations are given in \cite{carac}
and \cite{theop}.
\subsection{{\sc superstructure}  and {\sc autostructure}} 
The configuration interaction (CI) codes {\sc superstructure} (SS) \cite{ss}
and {\sc autostructure} (AS) \cite{as1,as2} use expansions
\be \Psi=\sum_m\Phi_m \ee
where the $\Phi_m$ are one-configuration functions.
The code SS is used only for calculating energies of bound states
and radiative transition probabilities: AS, which was developed from SS,
includes calculations for auto-ionizing states, auto-ionization
probabilities, and cross-sections for photo-ionization.

In using the CI codes for a level of a given configuration ${\cal C}$, we 
usually attempt to include in the expansion (6) at least all of the
states belonging to the complex of which ${\cal C}$ is a member.
\subsection{Methods used in OPAL}
OPAL uses single-configuration wave functions,
\be \Psi=\Phi_m, \ee
with one-electron orbitals calculated using potentials adjusted
empirically such as to give best agreement with experimental
energy-level data \cite{riw,irw}.

\section{Spectrum lines}
Inclusion of contributions from large numbers  of spectrum 
lines  is of crucial importance. 
\subsection{Inclusion of lines}
Our procedure is, first,
to calculate the monochromatic opacity cross-section  $\sigma(\nu)$
including only
continuum processes (photo-ionization and -detachment,
free-free processes and scattering) and to determine a mean
back-ground $\sigma_{\rm B}(\nu)$ such that, 
at each frequency point $\nu$, $\sigma(\nu)\ge\sigma_{\rm B}(\nu)$.
Let $\sigma_{\rm L}(\nu)$ be the contribution from a spectral
line. We define a quantity {\sc testl} and, at each frequency
mesh-point, include the line contribution if 
\be \sigma_{\rm L}(\nu)\ge\mbox{\sc testl}\times\sigma_{\rm B}(\nu).\ee
We  take {\sc testl} to be small (usually {\sc testl}$=10^{-8}$):
the contribution from each line may be small but the number of
lines may be very large.
\subsection{Fine structure}
Inclusion of the fine structures of
the spectrum lines can lead to a re-distribution of oscillator
strength which can give a significant increase in $\sigma_{\rm R}$
defined by (\ref{ross})       .
In the original OP work fine-structure was allowed for by 
methods described in section 4.5 of \cite{symp}: simple LSJ-coupling
formulae were used together with empirical estimates of averaged
spin--orbit parameters so as to determine level splittings.
It was checked that the method was very
stable, in that variation of the parameters over a wide range
gave hardly any change in the calculated values of $\sigma_{\rm R}$.
However, the method did not allow for the inclusion
of intercombination (spin-forbidden) lines which can be
important for highly ionized systems.
\subsection{Line profiles}
The quantity $\sigma_{\rm R}$ can be sensitive to the widths of the
spectrum lines, which are due to: radiation damping; thermal
Doppler effects; and pressure broadening. Both OP and OPAL use
similar empirical formulae for the pressure widths based on a
theory originally proposed by Griem \cite{griem}. They differ in
that OPAL adopts parameters determined from available experimental
data \cite{kandd} while OP uses results from ab-initio calculations
\cite{p1,p2,p3}. In the fairly small number of cases for
which direct comparisons have been made \cite{p2} the agreement between
OP and OPAL pressure-widths is very close.

It may be noted that $\sigma_{\rm R}$ is sensitive to pressure
broadening only for intermediate values of the density: at 
sufficiently low densities the pressure-widths are small
compared with those due to radiation damping and Doppler
effects; while at high densities the lines are blended to
form a quasi-continuum.
\section{Hydrogen and Helium}
The basic physical data for H and He (energy levels, radiative
transition probabilities, and cross-sections for photo-detachment,
photo-ionization and free--free transitions) are
known accurately and theories of pressure-broadening of
lines for those elements are well developed and should be
quite reliable. Fig. 2 gives OP and OPAL results for a H/He
mixture with mass-factions of $X=0.7$ for H and $Y=0.3$ for He. As is
to be expected, the agreement between OP and OPAL is generally very close
but there is a region with some differences for
$\log(R)=-1$ within the vicinity of $\log(T)=6$.
We discuss that region further.
\subsection{Hydrogen}\label{H}
The top part of Fig. 3 shows $\log(\kappa_{\rm R})$ for
pure H with $\log(R)=-1$ and $\log(T)=5$ to 7 and the lower 
part shows the first derivative,
\be \partial\log(\kr)/\partial\log(T), \label{parti}\ee
calculated in the approximation of using first differences.
The largest difference between OPand OPAL occurs for $\log(T)\simeq 6.0$ and
in that region there is a marked change in the behaviour 
of $\partial\log(\kr)/\partial\log(T)$, which becomes practically constant for 
$\log(T)>6.5$. There are three main contributions to hydrogen 
opacities in the region considered: (1) scattering of photons
by free electrons; (2) electron--proton free--free transitions;
(3) hydrogen bound--free transitions. Contributions from (3) become very 
small for $\log(T)$ significantly larger than 6.0, as H becomes
fully ionized.

OP calculations are first made on a mesh of values of $(\log(T),\log(\Ne))$
and interpolations to required values of $(\log(T),\log(\rho))$ are
then made using the code {\sc opfit.f} \cite{opfit}. Table 2 gives
results for the Hydrogen ionization equilibria at $\log(T)=6.0$:
values of $\log(R)$; values of $\log(\Ne)$; $\phi(1)$ (fraction
of neutral H); and $W(1)$ (H ground-state occupation probability)
from OP. The table also includes values of $\log(\Ne)$ and $\phi(1)$ 
from OPAL (data
kindly supplied by Dr C. A. Iglesias). The trends for OP may be
described as follows: with increasing density, $\phi(1)$ initially
increases due to the factor $\Ne/\Ue$ in (3) (pressure recombination)
but for the larger values of density $W(1)$ becomes small leading
to smaller values of $\phi(1)$ (pressure ionization). The values of 
$\phi(1)$ from OPAL are much larger than those from OPAL and
the additional amount of neutral H explains the difference between
the OP and OPAL opacities. 

It will be noted that at the lower densities of Table 2 the values of
$\phi(1)$ from OPAL continue to be larger than those from OP.
The reason is that the OPAL values of the $W(n)$ for $n>1$ are
larger than the OP values. However, those differences in ionization
equilibria at the lower densities do not have much effect on
the Rosseland-mean opacities.
\subsection{Helium}\label{He}
Fig. 4, for Helium, shows results similar to those of Fig. 3,
for Hydrogen. At $\log(R)=-1$ the maximum difference between the OP and OPAL
opacities occurs for $\log(T)\simeq 6.4$. That is the region where
the OP occupation probability for the He$^+$ ground-state becomes
small. The OPAL populations for He$^+$ will be much larger than 
those from OP.
\subsection{Smoothness}\label{smooth}
A further feature of Figs 3 and 4 may be noted. Both OP and
OPAL give rise to somewhat irregular appearances for the derivatives
$\partial\log(\kr)/\partial\log(T)$. Dr Iglesias informs us that when producing
a pure H or He table for OPAL it is necessary to do some
interpolating, which might explain the irregularities in the OPAL
data. The OP
work also involves interpolations. 
A check run was made for H using intervals of $\delta\log(T)=0.025$
and $\delta\log(N_{\rm e})=0.25$, that is to say one half of those
normally used, and it was found
to give close agreement with the results presented in Fig.~3.
\subsection{Summary on H and He}
The differences between OP and OPAL for level populations 
(and, more generally, for equations of state, EOS) have
been discussed in a number of previous papers (\cite{dgh}, \cite{IR96},
\cite{bs03}) and will not be discussed further here. Our present
concern is only with effects on opacities. For pure H and pure He
the largest differences in opacities, at $\log(R)=-1$, are 13\%
for H at $\log(T)=6.0$ and 20\% for He at $\log(T)=6.4$. For
smaller values of $\log(R)$ the differences are much smaller
(see Fig. 2). When other elements are included 
(the `metals') the differences between OP and OPAL are generally
much smaller, because in the regions concerned the contributions
to opacities from those other elements will generally be much
larger than those from H and He. Thus Fig. 1 shows
no sign of the discrepancy between OP and OPAL shown on Fig. 2.
\section{Carbon and Oxygen}
For the C and O RM calculations we include in (\ref{rm}) all states $\psi_p$ which 
belong 
to the ground-complex of the $(N-1)$-electron system, that is to say all states
with the same set of principal quantum numbers as the ground state. Thus
the system with $N=6$ electrons has a ground configuration of 
1s$^2$2s$^2$2p$^2$ and the ground-complex of the $(N-1)$-electron system 
contains the configurations 1s$^2$2s$^2$2p, 1s$^2$2s2p$^2$ and
1s$^2$2p$^3$. With those choices for the $\psi_p$, all
bound states of C and O can be represented
by expansions of the type (\ref{rm}). R-matrix calculations were made
for all bound states with an outer electron having quantum
numbers $n\ell$ with $n\le 10$ and $\ell\le$ {\sc lmax} with
{\sc lmax} varying between 2 and 4: states with $\ell>${\sc lmax}
were treated in hydrogenic approximations.

Independent evaluations of OP oscillator strengths for
C, N and O have
been made in \cite{nist}. The OP values are found to compare favourably
with experimental data and with data from other refined calculations.
For most transitions in C, N and O the OP data are recommended in \cite{nist}
as the
best available. For C and O, in addition to the R-matrix data,  we now
include data for inner-shell transitions from \cite{bs03}.

We recall that the OPAL website \cite{web} provides Rosseland-mean opacities
for `metal' mass-fractions \mbox{$Z\le 0.1$}. Figs 5 and 6 compare
OP and OPAL opacities for H/C and H/O mixtures with $X=0.9$ and 
$Z=0.1$. Agreement is fairly close. 

For the lower temperatures, $\log(T)\lesssim 5.5$, the
differences between OP and OPAL may be due to differences
in atomic data, with the OP RM data probably being the
more accurate. The maxima at higher temperatures 
($\log(T)\simeq 6.0$ for C and  6.2 for O)
are due to transitions with K-shell initial states (electrons
with principal quantum number $n=1$). For the higher densities
in those regions, $\log(R)\gtrsim -2$, the
opacities from OP are larger than those from OPAL, which is
an EOS effect. 

In the regions concerned the dominant ionisation stages for C and 
O are H-like and He-like and for OP dissolution occurs for outer
electrons having $n\gtrsim 3$, 
%electrons having $n$ greater than or equal to about 3,
while for OPAL it occurs at considerably  
larger values of $n$; hence OPAL will have more electrons in 
highly excited states, and less in states with $n=1$, giving smaller
opacities. 
For the reasons given in section 3.2 we consider that
OP may give the more accurate results for
the positions at which dissolution occurs.
\section{Sulphur}
For ions with numbers of electrons $N\le 11$ the ground-complexes 
for the $(N-1)$-electron cores contain only 1s, 2s and 2p electrons
but for $N>11$ 3d electrons must also be included and the
number of configurations in the complexes can become very large.
For such values of $N$ that number was, in general, too 
large for the inclusion of all such states in the RM expansions
and some further approximations were required. Thus even
for the case of $N=13$ the states 3s$^2$, 3s3p ands 3p$^2$ were
included but 3p3d and 3d$^2$ were omitted. For the case of S,
with $N\le 16$, such omissions occurred only for the earlier
ionization stages where the omitted core states would be rather
high and their omission would be unlikely to lead to serious error.

Fig. 7 compares OP and OPAL opacities for an H/S mix with $X=0.9$ and $Z=0.1$.
The level of accord is similar to that observed for C and O.
For S, $\kr$(OP) is larger than $\kr$(OPAL) at the higher densities in the 
K-shell region at $\log(T)\simeq 6.7$.  That is an EOS effect.
\section{Iron}
Iron plays a special r\^{o}le in calculations of opacities for
typical cosmic mixtures, in consequence of its comparatively high
abundance and the very large numbers of lines in its spectra. 
As the nuclear charge $Z$ increases all levels in a complex
eventually become degenerate (in the limit of $Z\rightarrow \infty$
the one-electron energies, when relativistic effects 
are neglected, depend only on the principal
quantum numbers $n$). With increasing $Z$ the levels in a complex move
downward and the number of bound levels in a complex 
therefore  increases. Consider iron with $N=$16 (Fe XI) which
has a ground configuration 3s$^2$3p$^4$ and a ground-complex
containing all configurations of the type 3s$^x$3p$^y$3d$^z$
with $x+y+z=6$.  Runs with AS give the number of spectroscopic
terms which belong to that complex, and which have energies
below the ionization limit, to be equal to 721. There are large 
numbers of radiative transitions between such states (the 
`$3\rightarrow 3$' transitions) and from such states
(the `$3\rightarrow n$' transitions with $n>3$).
It was first shown in the OPAL work that such transitions in 
iron ions with $N=14$ to 19 give rise to an important feature
in the Rosseland-mean, at $\log(T)\simeq 5.2$,which has come to be known as the 
`$Z$-bump' \cite{bump} (see Fig. 1).
\subsection{Summary of iron data used}
Calculations have been made in both 
LS coupling and in intermediate coupling (IC).
Table 3 gives a summary of the atomic data for iron used in 
various stages of the OP work. 

\subsubsection{Data from Kurucz}
For the first few ionization stages, $N=21$ to  26, data from Kurucz were used
\cite{K}, for nearly 7 million lines. The data were computed using
the code of Cowan \cite{cowan}. The contributions from
those stages is not normally of major importance for calculation
of the Rosseland means but can be of crucial importance for
radiative accelerations \cite{acc}.

\subsubsection{R-matrix calculations, RM}
RM calculations for energy-levels, oscillator strengths and
photo-ionization cross-sections were made for all ionization stages of iron
but for $N=21$ to 26 only the photo-ionization data were used
(much larger amounts of data being available from Kurucz).
In Table 3 the numbers of levels and lines are given
separately for $N=21$ to 28, $N=14$ to 20 (the region of the $Z$-bump) and for
$N=2$ to 13. For $N=2$ to 13 the data from RM are probably the
best available.

\subsubsection{Use of {\sc superstructure}}
The RM method was not capable of providing the
very large amounts of data in the region of the $Z$-bump, and
further data for over 3 million lines were calculated using 
SS \cite{plus}, and used in \cite{symp}.

\subsubsection{Inner-shell data from {\sc autostructure}}
More recently, inner-shell data from AS, for iron and other elements, have
been included \cite{bs03}. In \cite{bs03} it was found that, for
inner-shell data, use
of IC, in place of LS coupling, did not give any significant change in 
Rosseland means. Here we consider only LS coupling inner-shell data.

\subsubsection{Further data from {\sc autostructure}}
In the present paper we consider the replacement of data from
RM and SS, in the regions of the $Z$-bump ($N=14$ to 19), with
further outer-shell data from AS. Calculations were made using
both LS coupling and IC with over 30 million lines for the latter case.

\subsubsection{Photo-ionization}
The numbers given under the heading `PI' in Table 3 correspond
to the numbers of initial states for which photo-ionization
cross-sections were calculated: each cross-section contains
data for all available final states.
\subsection{Use of new AS data in LS coupling}
For ionization stages giving the $Z$-bump, $N=14$ to 19, we
tried replacing all old (RM plus SS) data with the new AS
data in LS coupling. We obtained Rosseland means close to
those from the original OP work which
provided a good independent check.
The only significant differences
were in fairly restricted regions where photo-ionization 
had been estimated using SS data (the SS estimates were rather
crude -- see \cite{plus}).
\subsection{Intercombination lines}
We replaced all old (RM plus SS) data for $N=14$ to 19
with new AS data in IC. The differences with Rosseland-means
from the previous subsection (AS data in LS coupling) was almost
entirely due to inclusion of the intercombination lines in
AS. Fig. 8 shows the percentage increase in the mean which
results from inclusion of the intercombination lines (up to 18\%
for $\log(R)=-4$). The inclusion of those lines gives
improved agreement with OPAL near the maximum of the 
$Z$-bump. But, on a log--log scale, our final results for the 
6-element mix do not appear greatly different to the eye from 
those shown in Fig. 1.
\subsection{Many-electron jumps}
Selection rules state that radiative transitions can occur only
between configurations which differ in the states of one electron.
In configuration-interaction calculations the states may be
given configuration labels but transitions can occur between
states having configuration labels differing by more than
one electron (the `many-electron jumps'). The effect can be
a further re-distribution of oscillator strength. We have made
runs in which the many electron jumps are omitted. Fig.
9 shows the percentage increase which results from inclusion
of those jumps. The effect is not large (no more than 8\%).
It is not included by OPAL.
\subsection{Results for an iron-rich mixture}
It has already been noted that the OPAL website \cite{web}
cannot provide opacities for mixtures with $Z>0.1$. A further
restriction is that it cannot provide data for cases in
which $Z$ is pure iron (that restriction arises because such cases can
be very sensitive to behaviours near minima in the iron
monochromatic opacities). About the most iron-rich case for
which we can obtain OPAL data is $X=0.9$, $Z=0.1$ and $Z$
containing C and Fe in the ratio of 2:1 by number fraction.

Fig. 10 compares  OP with OPAL for the iron-rich mixture.

At low densities inclusion of iron intercombination transitions
in the OP work lead to an increase in $\kr$ in the region of
the maximum of the $Z$-bump, giving close agreement between
OP and OPAL in that region. But the position of the $Z$-bump
given by OP is shifted to slightly higher temperatures,
compared with OPAL.

At higher densities there are some differences between the 
OP and OPAL results shown on Fig. 10. At those densities there
are some suggestions of rather irregular variations in the 
OPAL data which might be due to interpolation errors near
the edges of the domain of validity for the OPAL interpolations.
\section{The solar radiative interior}
Computed solar models are sensitive to radiative opacities
in the solar radiative interior, the region below the base
of the solar convective zone at $R_{\rm CZ}$. It is shown in two recent
papers (\cite{BA}, \cite{BSP}) that helioseismology provides
remarkably accurate measures of
$R_{\rm CZ}$, and hence stringent tests of the accuracy of solar
models. In those papers it is noted that recent
work leads to revisions in solar element abundances and that, 
when those revisions are taken into account, there are 
significant differences between values of $R_{\rm CZ}$ obtained
from helioseismology measurements and from solar models calculated
using OPAL opacities. 
In \cite{BA}, from a study using envelope models,
it is argued that in order to obtain a correct density
profile it is necessary to adopt opacities larger
than those from OPAL by an amount of 19\%; while in
\cite{BSP}, from a study of full evolution models, it
is found that an increase by 7\% is required.

At $R_{\rm CZ}$ the best estimates \cite{BSP} of the temperature 
and density are $\log(T)=6.338$ and $\log(\rho)=-0.735$ 
giving $\log(R)=-1.75$. In Fig. 11 we show the percentage
differences, \mbox{(OP -- OPAL)}, in $\kr$ against $\log(T)$ 
for the 6-element mix at $\log(R)=-1.75$.   At $\log(T)=6.338$  we 
find $\kr$(OP) to be larger than $\kr$(OPAL) by 5.0\% 
which is fairly close to the value of 7\% required in
\cite{BSP}. We find it to be highly unlikely that the
OPAL opacities could be in error by as much as 19\% as
suggested in \cite{BA}.
\section{Summary}
There are two main steps in opacity calculations. 
\begin{enumerate}
\item {\bf The EOS problem}, determination of
the populations of all species in a plasma which can lead to
absorption of radiation.
\item {\bf The atomic physics problem}, obtaining the atomic data which
control the efficiencies of the the radiative processes.
\end{enumerate}
\subsection{The EOS problem}
The OPAL and OP approaches to the EOS problem are very different
(see Section 3).

We encounter two cases for which differences between OP and
OPAL opacities are due to different treatments of the EOS problem. 

The first is for $\log(R)\simeq -1$ and  $\log(T)\simeq 6.0$ 
for H and 6.4 for He (see Section 3). For those cases
OP gives the ground states of H$^0$ and He$^+$ to be 
undergoing dissolution, leading to pressure ionisation,
while OPAL has much larger populations for those states.
The biggest difference is for He at $\log(T)\simeq 6.4$ and 
$\log(R)=-1$, for which $\kr$(OPAL) is larger than $\kr$(OP)
by 20\%. We are not aware of any experimental or observational 
evidence in favour of one result or the other.

The second case concerns K-shell transitions in `metals' at the 
higher densities, where OPAL has larger populations in 
excited states, and hence less in states with $n=1$ giving
smaller opacities. That causes the 5\% difference between
$\kr$(OP) and $\kr$(OPAL) at conditions corresponding to those
at the base of the solar convection zone (see Section 11).
The helioseismology evidence tends to favour the results
from OP. 
\subsection{Atomic data}
It is essential to consider both the quantity and the quality of
the atomic data used.

The  OP work was started, a bit more than 20 years ago, with the
ambitious intention of computing all of the required atomic
data using the sophisticated R-matrix method, RM. The OPAL work was started,
at about the same time, with the less ambitious approach of
using single-configuration wave functions computed with the
aid of parametric potentials adjusted empirically so as to give
best agreement with experimental energies. The OPAL work quickly led to the
discovery of the `$Z$-bump', a feature at $\log(T)\simeq 5.2$
produced by a very large number of iron lines. OP was  unable to
compute data for such large numbers of lines using the RM method,
but was able to make supplementary computations using the code
{\sc superstructure} which includes  some allowance for 
configuration-interaction effects. The first published OP
opacities had substantial differences from OPAL only at
rather high temperature and densities and in \cite{IR96} it was
suggested that OP was missing some data for inner-shell transitions.
That suggestion is confirmed as correct in our recent paper \cite{bs03}.
The RM and SS codes are both unsuitable for handling the inner shell
processes and we therefore used the code {\sc autostructure}, AS.
In the present paper we also include large amount of outer-shell
data from AS.

\section{Conclusions}
Our main conclusion is that there is good general agreement between
OP and OPAL opacities for the 6 elements H, He, C, O, S and Fe. We expect to
obtain equally good agreement for other elements once work is
complete on the calculation of further inner-shell data.

There are some indications that the OPAL interpolation procedures
described in \cite{corr} may introduce some errors near the edges
of the domains of validity which are claimed.

At lower temperatures, $\log(T)\lesssim 5.5$, there are some modest
differences between OP and OPAL which may be a consequence of the
greater sophistication of the OP atomic-data work. At higher
temperatures and higher densities, $\log(R)\gtrsim -2$, the
OP opacities tend to be larger than those from OPAL in 
consequence of the use of different equations of state. There is
some helioseismology evidence that the OP results are the more
accurate.
\section*{Acknowledgments}
We thank Drs F. J. Rogers and C. A. Iglesias for their friendly
comments on the present work. We also thank Drs. H. M. Antia, J. N. Bahcall and
F. Delahaye  for their helpful comments.
%
%\clearpage
%
%  BIBLIOGRAPHY
% 

%
\bsp
%
% TABLES
%
\clearpage
\begin{table}
\begin{center}
\caption{The 6-element mixture of references \protect\cite{IR96} and \protect\cite{bs03}.}
\begin{tabular}{lll} 
\hline
Element & Number fraction \\
\hline
H    &$ 9.071(-1)^\ast$\\
He   &$ 9.137(-2)$\\
C    &$ 4.859(-4)$\\
O    &$ 9.503(-4)$\\
S    &$ 9.526(-5)$\\
Fe   &$ 3.632(-5)$\\
\hline
\multicolumn{2}{c}{$^\ast 9.071(-1)=9.071\times 10^{-1}.$}
\end{tabular}
\end{center}
\end{table}
\begin{table}
\begin{center}
\caption{Hydrogen ionization equilibrium at $\log(T)=6.0$.}
\begin{tabular}{ccccccc}
\hline
 & \multicolumn{3}{c}{OP} &\,\, 
& \multicolumn{2}{c}{OPAL}\\ \cline{2-4}
\cline{6-7}\\
$\log(R)$  & $\log(\Ne)$ & $\phi(1)$ & $W(1)$ & \,\, & $\log(\Ne)$ & $\phi(1)$
\\
\hline
$-2.776 $&$ 21.00 $&$ 5.304(-4)$&$ 0.960 $&$ $&$ 21.00 $&$ 1.090(-2)$\\
$-2.276 $&$ 21.50 $&$ 1.324(-3)$&$ 0.863 $&$ $&$ 21.49 $&$ 1.506(-2)$\\
$-1.775 $&$ 22.00 $&$ 2.689(-3)$&$ 0.575 $&$ $&$ 22.00 $&$ 2.101(-2)$\\
$-1.276 $&$ 22.50 $&$ 2.751(-3)$&$ 0.190 $&$ $&$ 22.49 $&$ 2.546(-2)$\\
$-0.776 $&$ 23.00 $&$ 1.491(-3)$&$ 0.034 $&$ $&$ 22.98 $&$ 4.164(-2)$\\
$-0.276 $&$ 23.50 $&$ 6.424(-4)$&$ 0.005 $&$ $&$ 23.45 $&$ 1.069(-1)$\\
\hline
\end{tabular}\end{center}\end{table}
\begin{table}
\begin{center}
\caption{Atomic data for iron.}
\begin{tabular}{rccrrrrrr}
\hline
$N$ & Method & Coupling & Levels & \,\, & Lines & & PI\\
\hline
21--26 & Kurucz & IC & 65\,\,427 & \,\, & 6\,\,920\,\,198 & &---\\
20--26 & RM     & LS & 8\,\,925     & & 368\,\,445         & & 4\,\,479\\
14--20 & RM & LS & 6\,\,502 &\,\,& 276\,\,450 & &4\,\,639\\
2--13  & RM & LS & 4\,\,232 &\,\,& 151\,\,974 & \,\, & 3\,\,559\\
14--19 & SS & LS & 134\,\,635 &\,\,& 3\,\,295\,\,773 & &2\,\,537 \\
14--19 & AS, outer & LS & 371\,\,483 &\,\,& 4\,\,594\,\,253 &\,\, & 8\,\,006\\
14--19 & AS, outer & IC & 1\,\,011\,\,266 &\,\,& 30\,\,555\,\,846 & \,\, & 
20\,\,851\\
2--14  & AS, inner & LS & (7\,\,688)$^\ast$ &\,\, &2\,\,145\,\,442 & &1\,\,339\\
\hline 
\multicolumn{2}{c}{$^\ast$ Lower (initial) levels only.}
\end{tabular}\end{center}\end{table}
%
% FIGURES 
%
\clearpage
\begin{center}
\epsfig{file=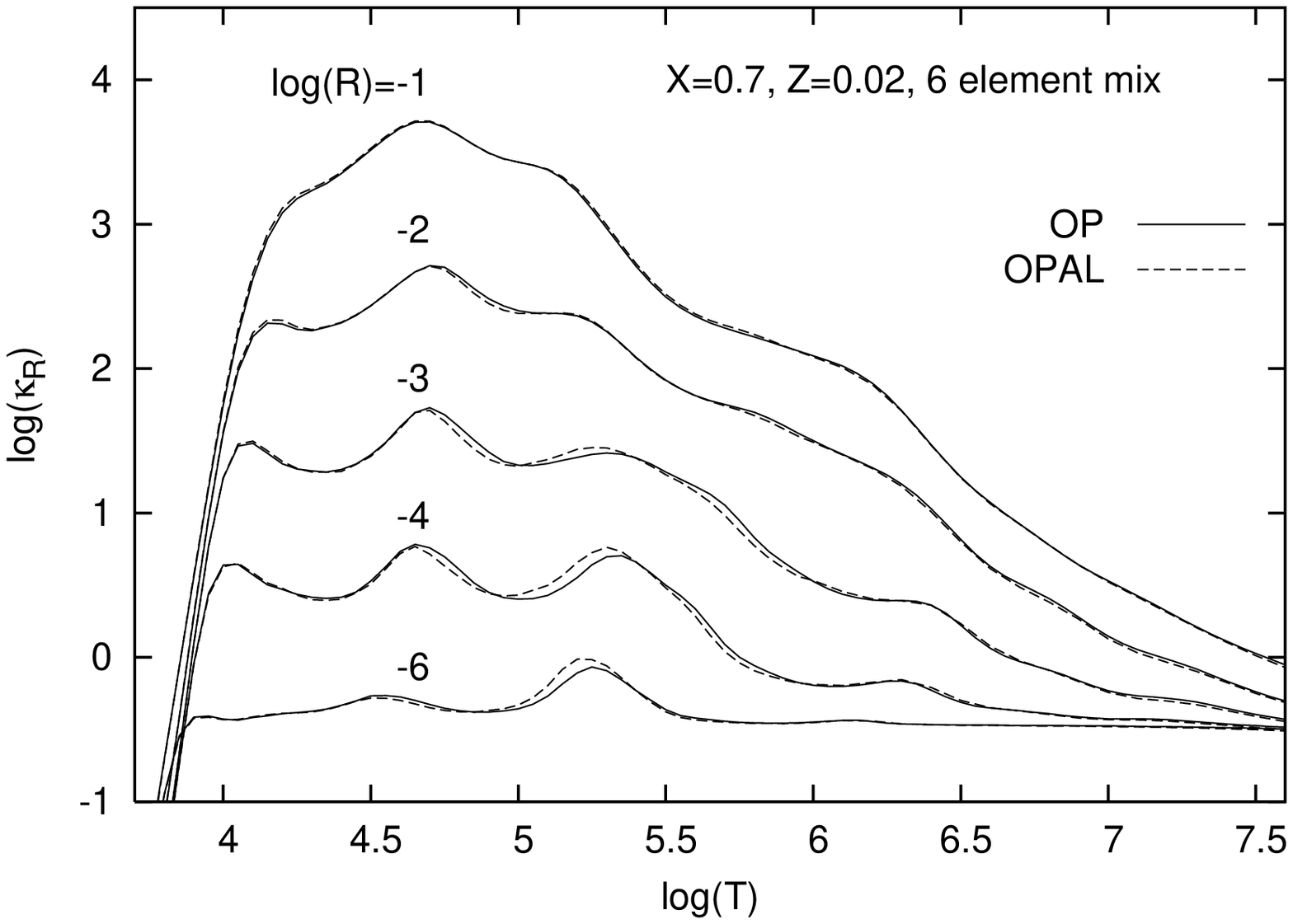,angle=270,width=10cm}
\end{center}
\vspace{2mm}
{\bf Figure 1.} Comparisons of $\log(\kr)$  from OP and OPAL for the 6-element
mixture of Table 1. OP from \cite{bs03}, OPAL from \cite{web}.
Curves are labelled by values of $\log(R)$ where $R=\rho/T_6^3$, $\rho$ is
mass density in g cm$^{-3}$ and $T_6$ is $10^6\times T$ with $T$ in K.
\vspace{10mm}
\begin{center}
\epsfig{file=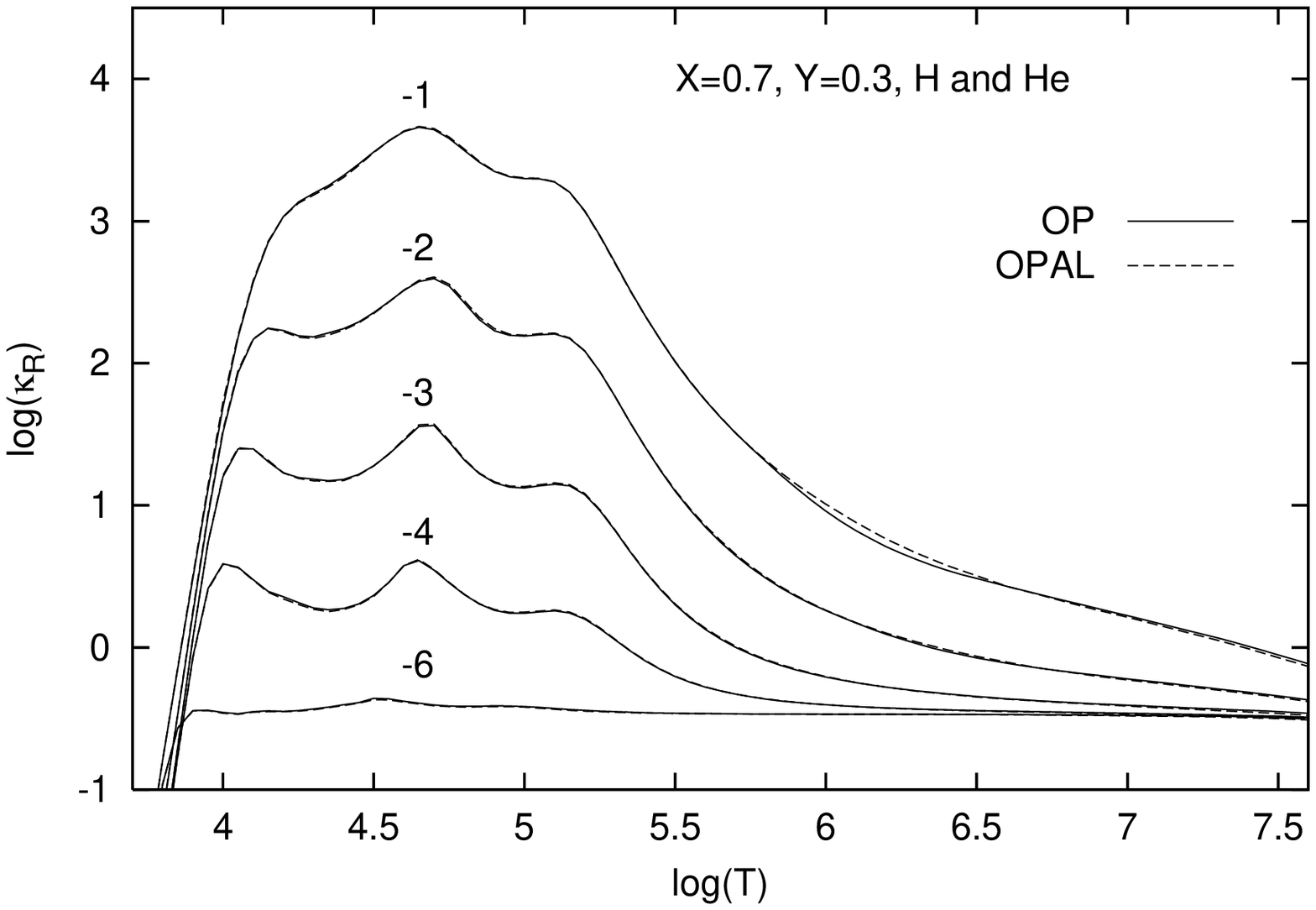,angle=270,width=10cm}
\end{center}
\vspace{2mm}
{\bf Figure 2.} Comparisons of $\log(\kr)$  from OP and OPAL for a
H/He mixture with mass fractions $X=0.7$ for H and $Y=0.3$ for He. 
Curves are labelled by values of $\log(R)$.
\clearpage
\begin{center}
\epsfig{file=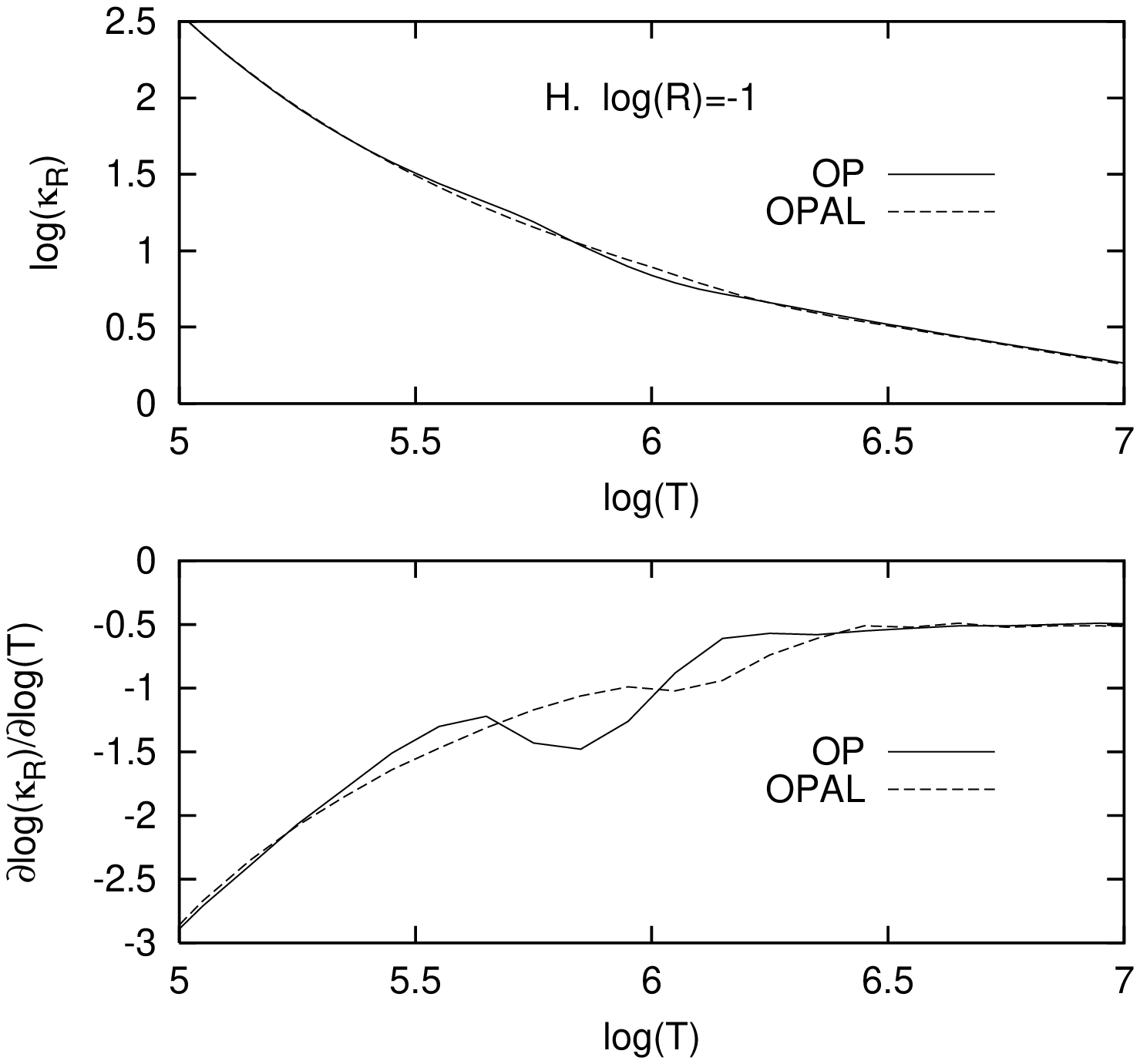,angle=270,width=10cm}
\end{center}
\vspace{2mm}
{\bf Figure 3.} Upper figure: $\log(\kr)$ for pure H, $\log(R)=-1$, 
$\log(T)=5$ to 7.
Lower figure: the derivative $\partial \log(\kr)/\partial \log(T)$
calculated using first differences.
\vspace{10mm}
\begin{center}
\epsfig{file=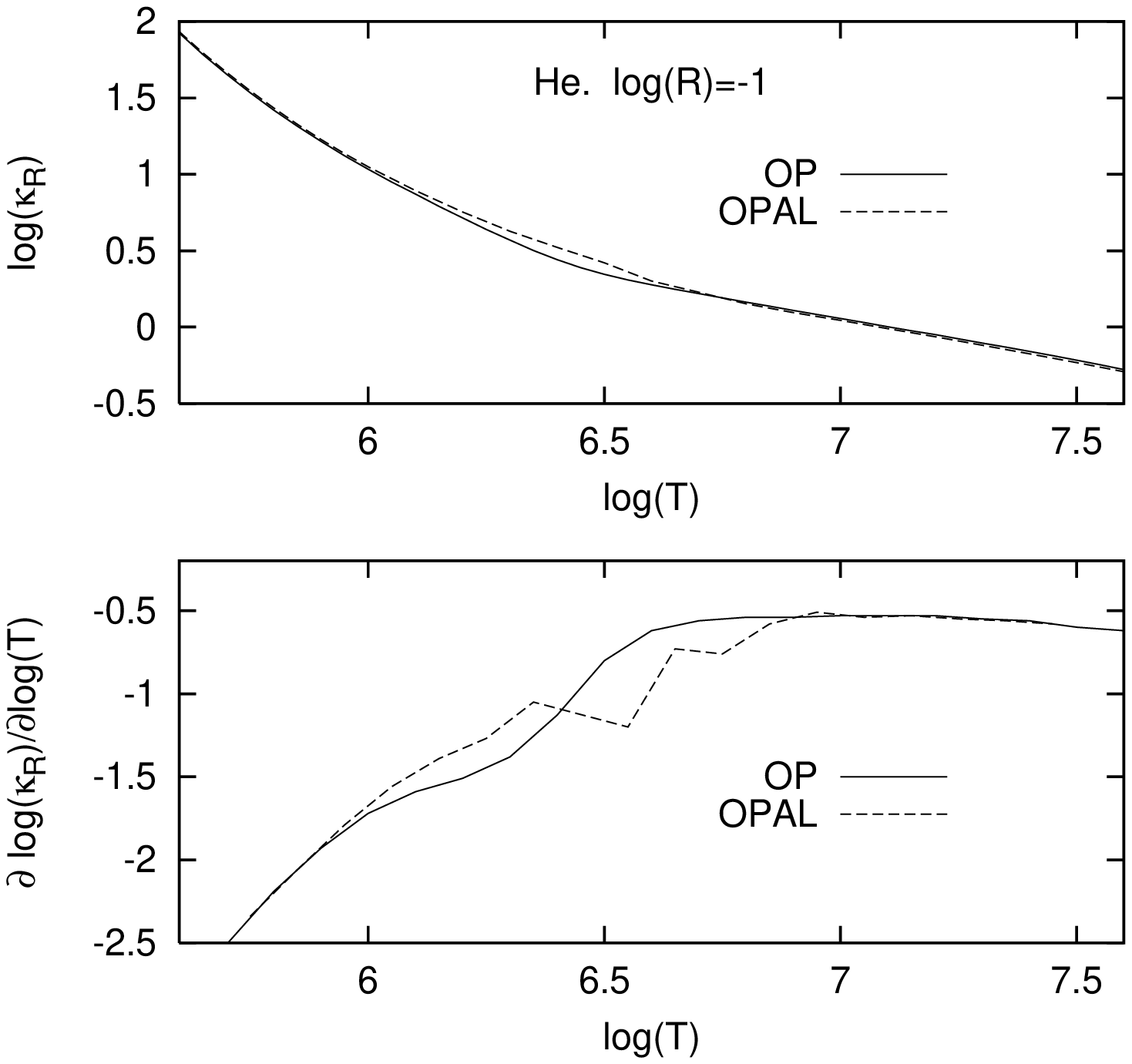,angle=270,width=10cm}
\end{center}
\vspace{2mm}
{\bf Figure 4.} Upper figure: $\log(\kr)$ for pure He, $\log(R)=-1$, 
$\log(T)=5.7$ to 7.5.
Lower figure: the derivative $\partial \log(\kr)/\partial \log(T)$
calculated using first differences.
\clearpage
\begin{center}
\epsfig{file=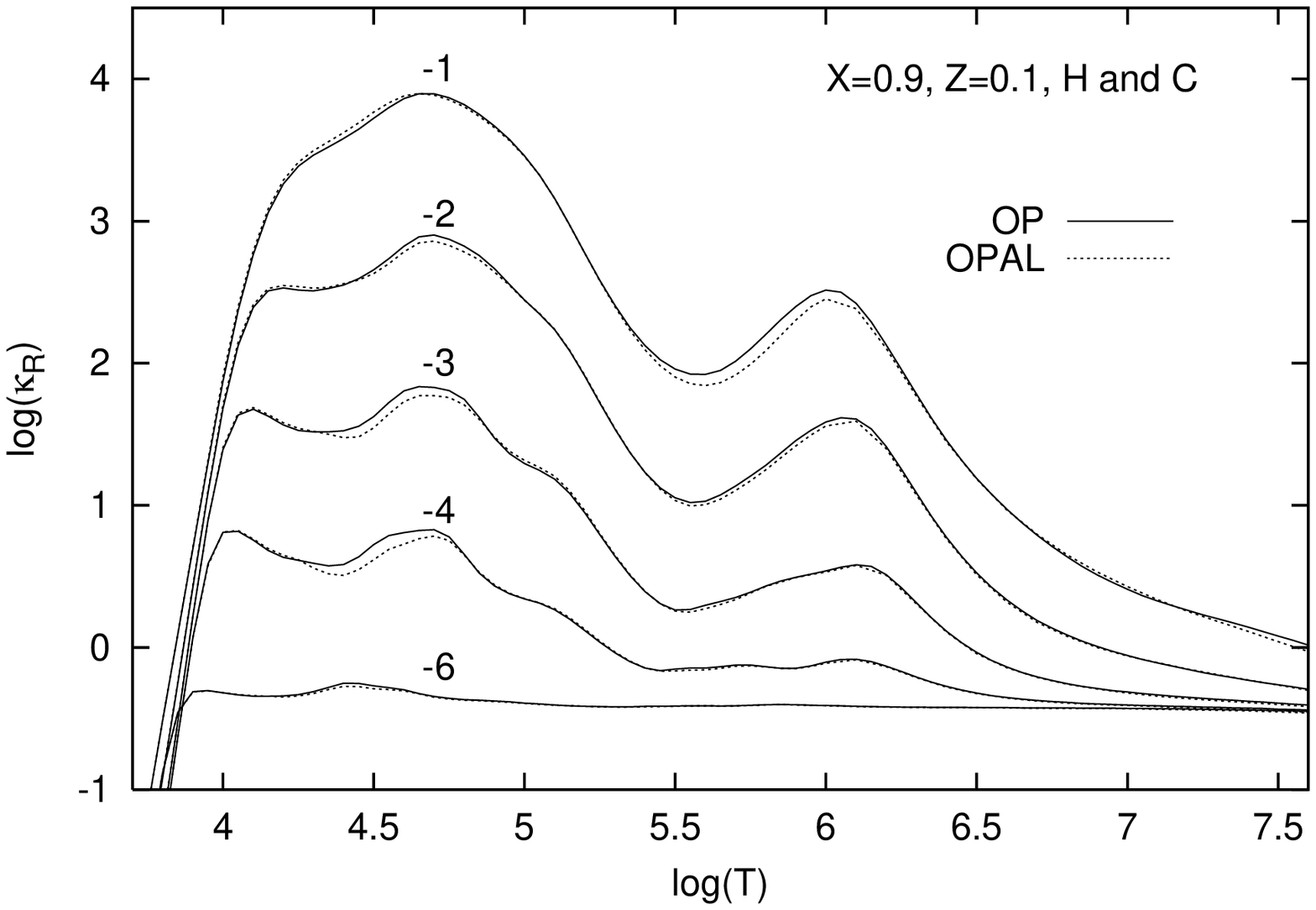,angle=270,width=10cm}
\end{center}
\vspace{2mm}
{\bf Figure 5.} Comparisons of $\log(\kr)$  from OP and OPAL for a
H/C mixture with mass fractions $X=0.9$ for H and $Z=0.1$ for C. 
Curves are labelled by values of $\log(R)$.
\vspace{10mm}
\begin{center}
\epsfig{file=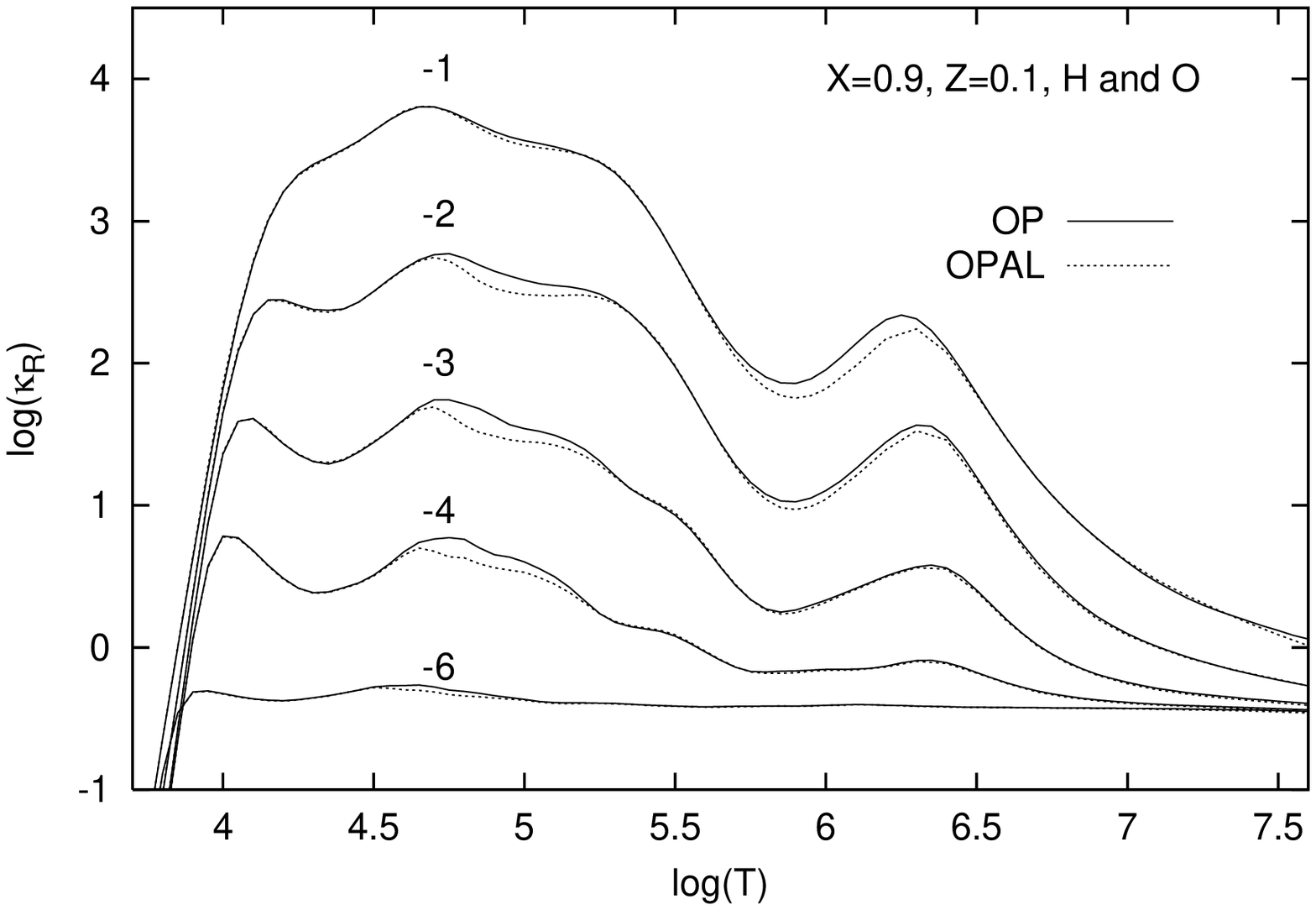,angle=270,width=10cm}
\end{center}
\vspace{2mm}
{\bf Figure 6.} Comparisons of $\log(\kr)$  from OP and OPAL for a
H/O mixture with mass fractions $X=0.9$ for H and $Z=0.1$ for O. 
Curves are labelled by values of $\log(R)$.
\clearpage
\begin{center}
\epsfig{file=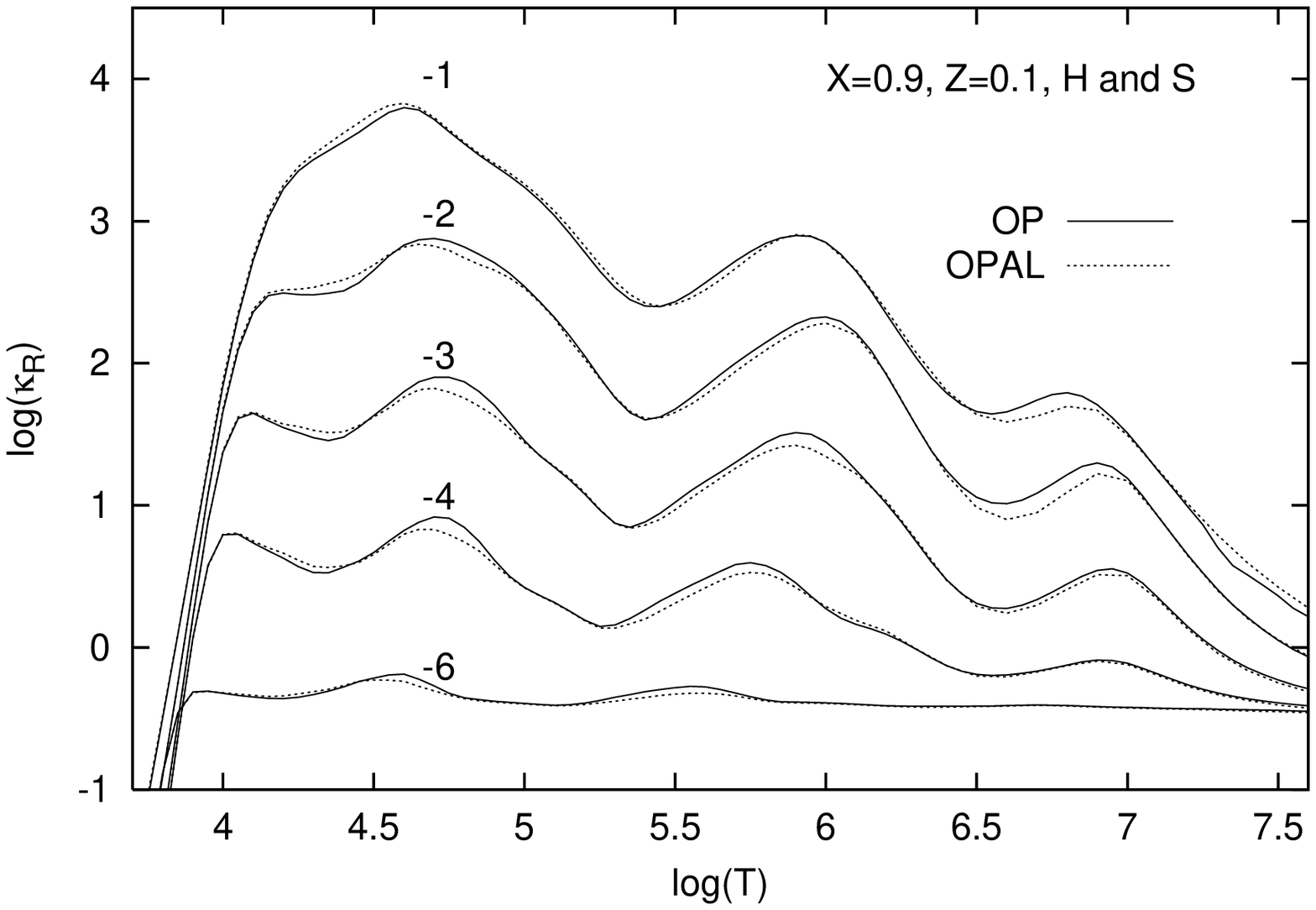,angle=270,width=10cm}
\end{center}
\vspace{2mm}
{\bf Figure 7.} Comparisons of $\log(\kr)$  from OP and OPAL for a
H/S mixture with mass fractions $X=0.9$ for H and $Z=0.1$ for S. 
Curves are labelled by values of $\log(R)$.
\vspace{10mm}
\begin{center}
\epsfig{file=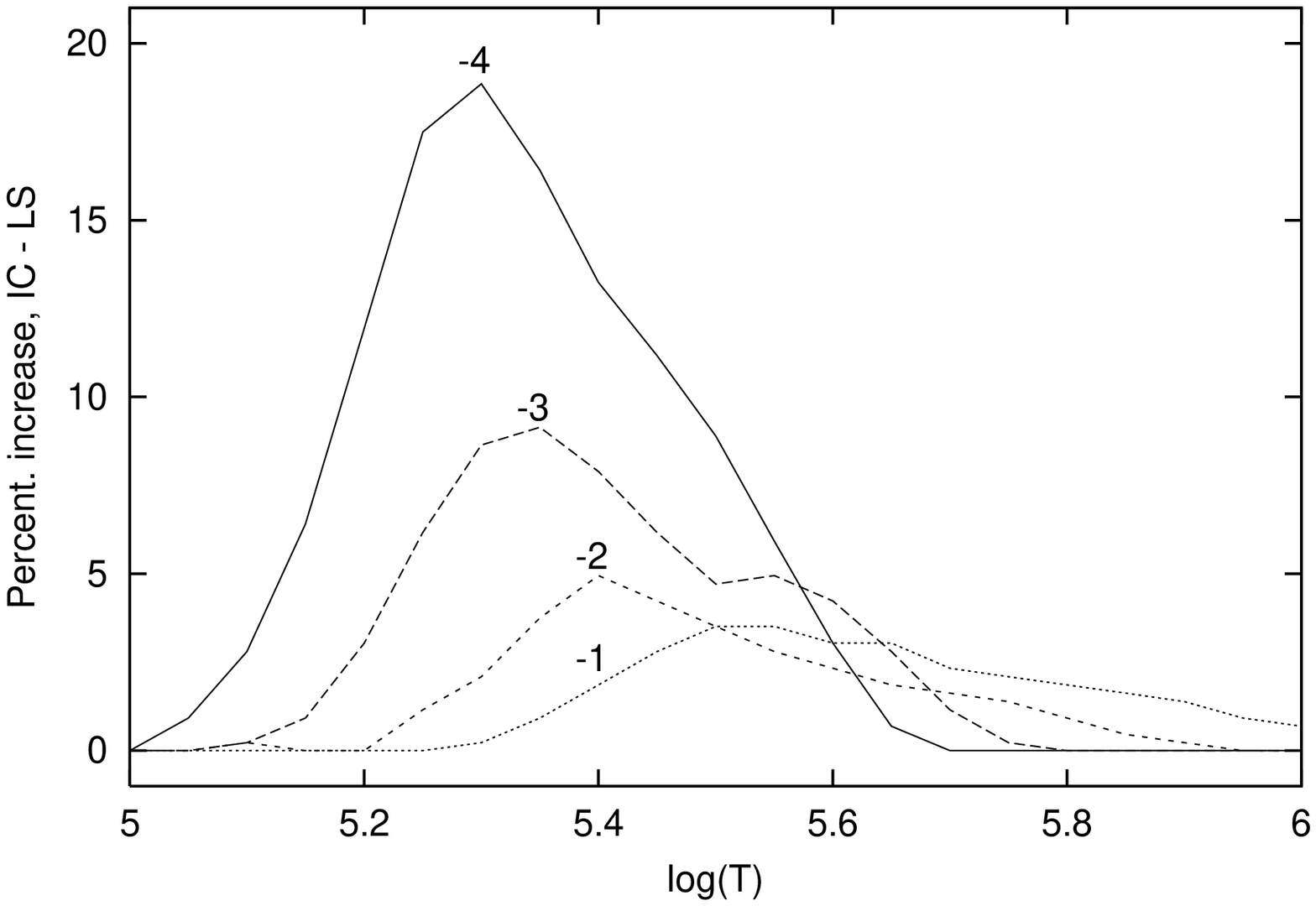,angle=270,width=10cm}
\end{center}
\vspace{2mm}
{\bf Figure 8.} Percentage increase in $\kr$, for the
6-element mix, which  results from the inclusion of iron intercombination 
lines. Curves are labelled by values of $\log(R)$.
\clearpage
\begin{center}
\epsfig{file=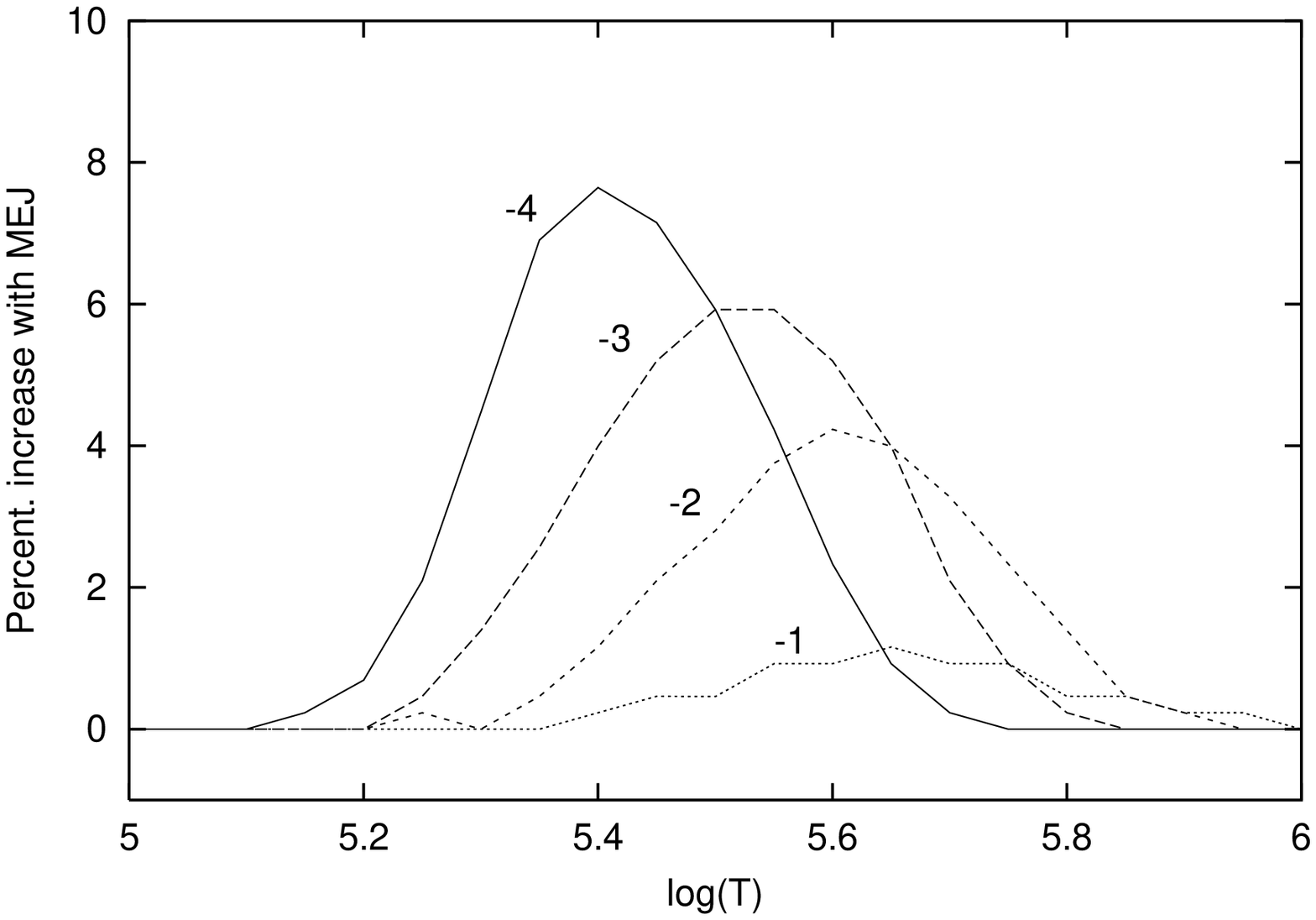,angle=270,width=10cm}
\end{center}
\vspace{2mm}
{\bf Figure 9.} Percentage increase in $\kr$, for the 6-element mix,
which results from the inclusion of iron many-electron-jumps (MEJ).
Curves are labelled by values of $\log(R)$.
\vspace{10mm}
\begin{center}
\epsfig{file=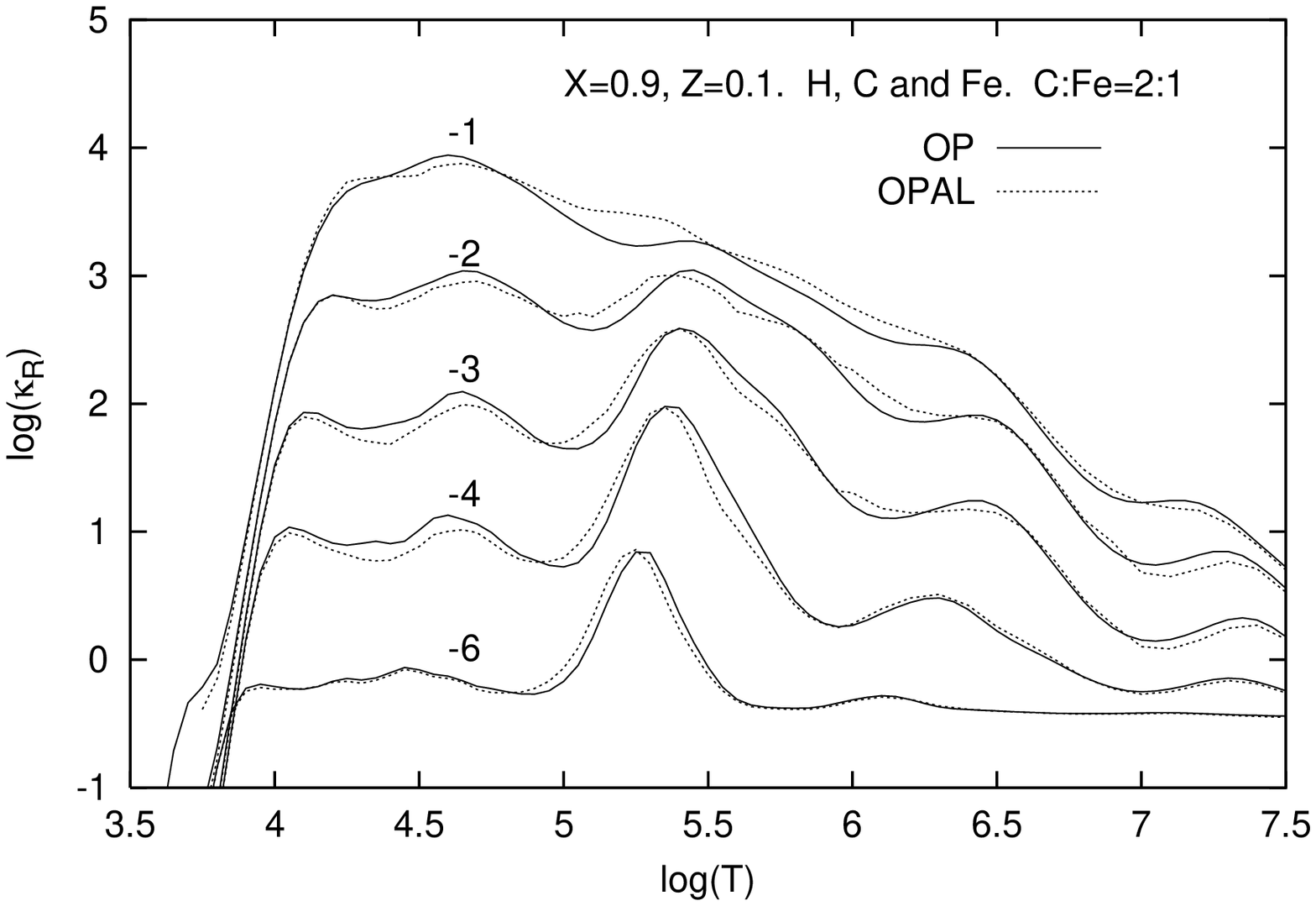,angle=270,width=10cm}
\end{center}
\vspace{2mm}
{\bf Figure 10.} Comparisons of $\log(\kr)$ from OP and OPAL for an iron-rich
mixture: $X=0.9$, $Z=0.1$ and C:Fe=2:1 by number fraction.
Curves are labelled by values of $\log(R)$.
\vspace{10mm}
\clearpage
\begin{center}
\epsfig{file=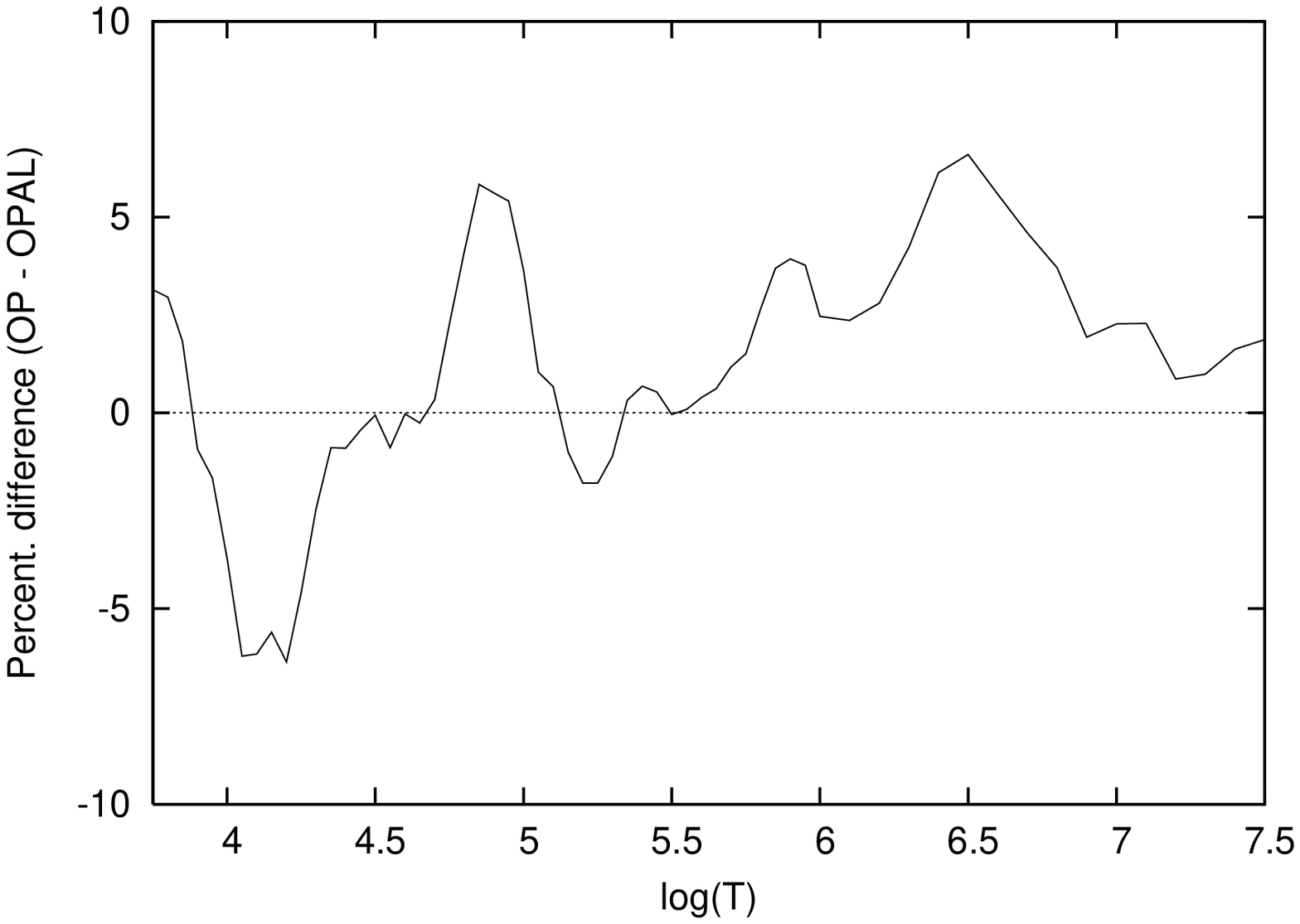,angle=270,width=10cm}
\end{center}
\vspace{2mm}
{\bf Figure 11.} Percentage differences, (OP -- OPAL), between  $\kr$ 
for the 6-element mix at $\log(R)=-1.5$.
\label{lastpage}

\begin{thebibliography}{99}
\bibitem{symp} \cite{symp}  Seaton M. J., Yu Yan, Mihalas D., Pradhan A. K., 1994, MNRAS,  266, 805
\bibitem{opal} \cite{opal}  Rogers F. J., Iglesias C. A., 1992, ApJS,  79, 507
\bibitem{IR96} \cite{IR96}  Iglesias C. A., Rogers F. J., 1996, ApJ,  464, 943
\bibitem{bs03} \cite{bs03}  Badnell N. R., Seaton M. J., 2003, J. Phys. B,  36, 4367
\bibitem{web}  \cite{web}   www-phys.llnl.gov/Research/OPAL/index.html
\bibitem{eos1} \cite{eos1}  Hummer D. G., Mihalas D., 1988, ApJ,  331, 794
\bibitem{adoc1}\cite{adoc1} Seaton M. J., 1987, J. Phys. B,  20, 6363
\bibitem{ghr}  \cite{ghr}   Graboske H. C., Harwood D. C., Rogers F. J., 1969, Phys. Rev.,  186, 210
\bibitem{corr} \cite{corr}  Rogers F. J., Iglesias C. A., 1993, ApJ,  401, 361
\bibitem{adoc2}\cite{adoc2} Berrington K. A., Burke P. G., Butler K., Seaton M. J.,
                	    	  Storey P. J., Taylor K. T., Yu Yan, 1987, J. Phys. B,  20, 6397
\bibitem{carac}\cite{carac} Seaton M. J., Zeippen C. J., Tully J. A., Pradhan A. K., Mendoza C.,
                	    	 Hibbert A., Berrington K. A., 1992, Rev. Mex. Astron. Astrofis.,  23, 19
\bibitem{theop}\cite{theop} The Opacity Project Team, The Opacity Project, IOP Publishing, Bristol
\bibitem{ss}   \cite{ss}    Eissner W., Jones M., Nussbaumer H., 1974, Comp. Phys. Comm., 8, 270
\bibitem{as1}  \cite{as1}   Badnell N. R., 1987, J. Phys. B,  19, 3827
\bibitem{as2}  \cite{as2}   Badnell N. R., 1997, J. Phys. B,  30, 1
\bibitem{riw}  \cite{riw}   Rogers F. J., Wilson B. G., Iglesias C. A., 1988, Phys. Rev. A, 38, 5007
\bibitem{irw}  \cite{irw}   Iglesias C. A., Rogers F. J., Wilson B. G., 1992, ApJ,  397, 717
\bibitem{griem}\cite{griem} Griem H. R., 1968, Phys. Rev.,  165, 258
\bibitem{kandd}\cite{kandd} Dimitrijevi\'{c} M. S., Konjevi\'{c} N. S., 1981, in Wende B., ed.,
                	    	  Spectral Line Shapes. de Gruyter, Berlin,  p. 211
\bibitem{p1}   \cite{p1}    Seaton M. J., 1987, J. Phys. B,  20, 6431
\bibitem{p2}   \cite{p2}    Seaton M. J., 1988, J. Phys. B,  21, 3033
\bibitem{p3}   \cite{p3}    Burke V. M., 1992, J. Phys. B,  25, 4917
\bibitem{opfit}\cite{opfit} Seaton M. J., MNRAS, 1993,  265, L25
\bibitem{dgh}  \cite{dgh}   Hummer D. G., 1988, AIP Conf. Proc.,  168, 1
\bibitem{nist} \cite{nist}  Wiese W. L., Fuhr J. R., Deters T. M., 1996,
                	    	 Atomic Transition Probabilities for Carbon, Nitrogen and Oxygen:
                	    	 A Critical Compilation, J. Phys. Chem. Ref. Data, Monograph 7
\bibitem{bump} \cite{bump}  Iglesias C. A., Rogers F. J., Wilson B. G., 1987, ApJ,  322, L45
\bibitem{K}    \cite{K}     Kurucz R. L., 1988, in McNally D., ed., Trans. IAU, XXB, Kluwer, Dordrecht, p. 168
\bibitem{cowan}\cite{cowan} Cowan R. D., 1981, The Theory of Atomic Structure and Spectra.
                	    	 Univ. California Press, Berkeley, CA
\bibitem{acc}  \cite{acc}   Seaton M. J., 1997, MNRAS,  289, 700
\bibitem{plus} \cite{plus}  Lynas-Gray A. E., Storey P. J., Seaton M. J., 1995, J. Phys. B,  28, 1995
\bibitem{BA}   \cite{BA}    Basu S., Antia H. M., 2004, ApJ, submitted (astro-ph/0403485)
\bibitem{BSP}  \cite{BSP}   Bahcall J. N., Serenelli A. M., Pinsonneault M., ApJ, submitted (astro-ph/0403604)
\bibitem{opale}\cite{opale} Rogers F. J., 1981, Phys. Rev. A, 23, 1008
\bibitem{nayf} \cite{nayf}  Nayfonov A., D\"{a}ppen W., Hummer D. G., Mihalas D., 1999, ApJ, 526, 451
\bibitem{RI92} \cite{RI92}  Rogers F. J., Iglesias C. A., 1992, Rev. Mex. Astron. Astrofis., 23, 133
\end{thebibliography}
\end{document}